# Super resolution of histopathological frozen sections via deep learning preserving tissue structure

Elad Yoshai[1], Gil Goldinger[2], Miki Haifler[2,3] and Natan T. Shaked[4,*]

[1] School of Electrical and Computer Engineering, Tel Aviv University, Tel Aviv, Israel
[2] Chaim Sheba Medical Center, Ramat Gan, Israel
[3] Sackler Faculty of Medicine, Tel Aviv University, Tel Aviv, Israel
[4] Department of Biomedical Engineering, Tel Aviv University, Tel Aviv, Israel
[*] Corresponding author: nshaked@tau.ac.il

## Abstract

**Histopathology plays a pivotal role in medical diagnostics. In contrast to preparing permanent sections for histopathology, a time-consuming process, preparing frozen sections is significantly faster and can be performed during surgery, where the sample scanning time should be optimized. Super-resolution techniques allow imaging the sample in lower magnification and sparing scanning time. In this paper, we present a new approach to super resolution for histopathological frozen sections, with focus on achieving better distortion measures, rather than pursuing photorealistic images that may compromise critical diagnostic information. Our deep-learning architecture focuses on learning the error between interpolated images and real images; thereby it generates high-resolution images while preserving critical image details, reducing the risk of diagnostic misinterpretation. This is done by leveraging the loss functions in the frequency domain, assigning higher weights to the reconstruction of complex, high-frequency components. In comparison to existing methods, we obtained significant improvements in terms of Structural Similarity Index (SSIM) and Peak Signal-to-Noise Ratio (PSNR), as well as indicated details that lost in the low-resolution frozen-section images, affecting the pathologist's clinical decisions. Our approach has a great potential in providing more-rapid frozen-section imaging, with less scanning, while preserving the high resolution in the imaged sample.**

## Introduction

Histopathology, a fundamental component of medical diagnostics, offers crucial insights into cellular-level tissue structures. In contrast to permanent sections, the preparation of which is a time-consuming process, frozen sections have emerged as an indispensable tool for enabling rapid intraoperative diagnosis, providing immediate evaluation of tissue samples during surgical procedures [1,2]. This approach involves the swift freezing of fresh tissue specimens, followed by sectioning for prompt microscopic examination, facilitating real-time decision-making for surgical interventions. Intraoperative diagnostic information is vital for surgeons to make well-informed decisions regarding tumor margins, extent of tissue excision, and overall treatment strategies. By offering immediate feedback on tissue pathology, frozen sections assist surgeons in achieving optimal surgical outcomes, minimizing the risk of incomplete tumor removal, and reducing the need for additional procedures.

Histopathological techniques may present limitations in image resolution and quality, potentially affecting diagnostic accuracy [3,4]. When processing time is an issue, especially in performing frozen sections during surgery, one might prefer to shorten scanning time by imaging it in lower magnification and resolution, allowing imaging a larger field of view at each scanning instance. Furthermore, scanning and stitching many images might cause more jittering problems and missing areas in the imaged sample [5].

Conventional super-resolution techniques have shown great promise in improving the quality and resolution of histological images. However, these techniques can also be time-consuming and computationally expensive [6], requiring significant digital processing power and specialized equipment. Deep learning-based super-resolution techniques offer an alternative approach that can save digital processing time [5], which can be critical in surgeries [2]. By using deep learning algorithms, it is possible to train models to learn the complex relationships between low-resolution and high-resolution images, enabling the generation of high-quality images from lower-quality inputs.

Recently, deep-learning-based super resolution has achieved promising results based on generative adversarial networks (GAN) [7] with architectures such as SRGAN, ESRGAN and Real-ESRGAN [8-12]. Moreover, in the emerging field of diffusion models [13,14], there are models such as SR3 [15] and SR3+ [16] that provide sharp and photo-realistic super-resolution images with upscale of even x8. However, these models frequently provide high perception results at the cost of low distortion measures accuracy [17], which measures the correspondence to the real high-resolution images.

The RCAN [18] architecture adds an attention mechanism [19] to select different features in the channels. DFCAN [20] uses for sparse-microscopy images a network architecture based on Fourier transform on the feature map with residual connection to rescale the feature map, and an SSIM-integrated loss. WA-SRGAN [21] uses wide-attention SRGAN on histological images and applies improved Wasserstein [22] with a gradient penalty to stabilize the GAN model while training. CARN [23] proposes a channel attention retention architecture to balance the efficiency and accuracy of the model. ESRT [24] suggests a light-weight backbone transformer [25] to improve the transformer's large memory occupation, computation, and inference time for the super-resolution task.

In the medical field, and specifically in histopathology where the diagnosis is done based on those images, it is critical to provide high accuracy super-resolution image in terms of distortion measures such as Structural Similarity Index (SSIM) [26] and Peak Signal-to-Noise Ratio (PSNR), rather than providing a sharp photo-realistic super resolution images with low correspondence to the real images, which may cause a wrong diagnosis.

In the field of histopathology, resolution increase by deep learning has been mainly performed on permanent sections, rather than on frozen sections. In the present paper, we propose a new architecture for obtaining super-resolution imaging of frozen sections, for the first time. For frozen sections, obtaining super-resolution to speed up sample scanning is more critical than for permanent sections since this procedure is used for rapid intraoperative diagnosis. Furthermore, since the rapid frozen section fixation process damages the tissue structure more than the permanent section process, it is more critical to produce reliable super-resolution images without high perception results for obtaining photorealistic images at the cost of low distortion measures. Our new approach is based on bilinear interpolation as a baseline that propagated through a residual connection an attention U-net [27,28] architecture that corrects the interpolation error. This makes the training process faster and more efficient, as the up-sampled image is used as a baseline, and the network is focused on correcting its error from the real image. The main idea is that as each pixel in the down-sampled image contains the average intensity of the related pixels in the super-resolution image, the model will be better using it in the residual feedforward as an initial baseline and put the effort of the network to learn only the error between the interpolated image and the super-resolution target.

Furthermore, as the high spatial frequencies in the image are more challenging to reconstruct, we used a new loss, weighted-frequency-loss, which utilizes the Fourier transform of the real and predicted images and the weighted difference between them, putting more penalty on the higher spatial frequencies.

## Methods

### Data collection and preprocessing

The data used to train our models was frozen-section pathology slides of kidney cancers, obtained from The Cancer Genome Atlas (TCGA), a publicly funded project initiated by the US National Cancer Institute and the US National Human Genome Research Institute. Whole-slide-images [29] in SVS format of full slides were downloaded and pre-processed into overlapping patches of 256x256-pixel RGB images with the highest resolution available. The data was shuffled, and totally, we obtained 45,050 256x256-pixel RGB frozen patches for training and 2,651 256x256-pixel RGB frozen patches for testing. We prepared two datasets, one with downsample of x4 and the other with downsample of x8, so for each of the datasets the target of the model is to reconstruct the original patch from the down sampled patch.

### Model architecture

Our model uses the bi-linear interpolation of the down-sampled input image as a residual feedforward and applies the attention U-net to correct the interpolation error, as can be seen in Figure 1. This enables fast training due to the interpolation feedforward baseline and leverages the attention mechanism in the attention U-net architecture to selectively focus on different parts of the input data and its feature maps.

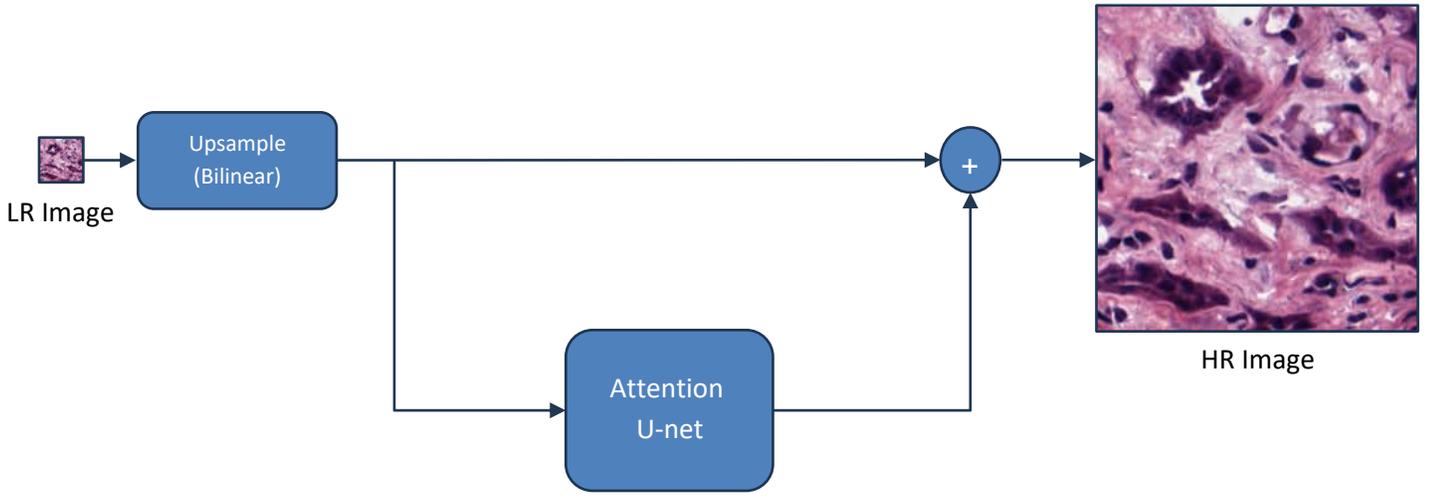

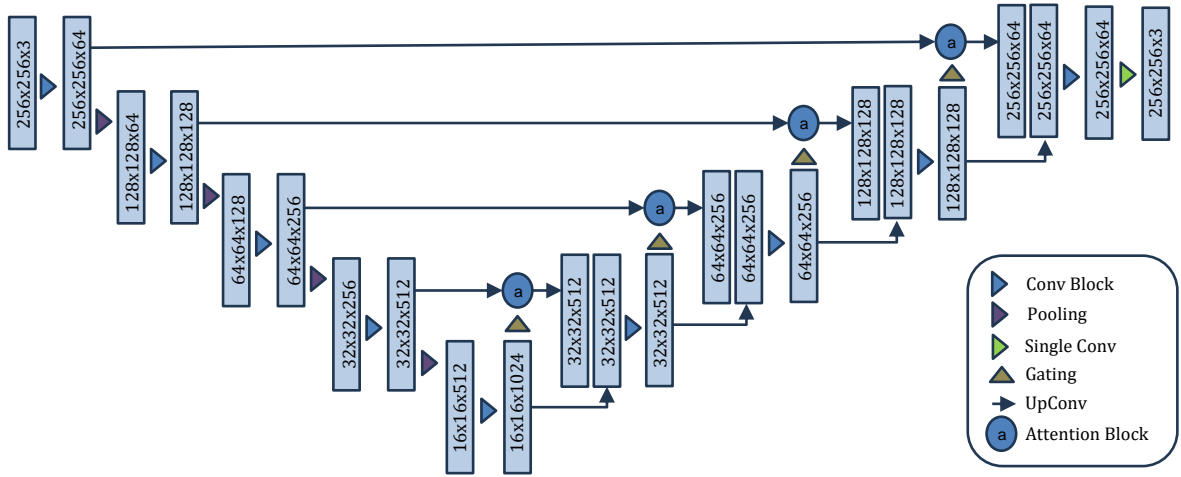

**Figure 1**. (a) Proposed network architecture with residual upsample as the baseline, and the attention U-net as a pixel-wise correction for faster training with better distortion measures (PSNR/SSIM). (b) The attention U-net architecture.

## Loss function

Since loss functions such as MSE and MAE rely on the mean error between the generated and the target images, it becomes challenging to reconstruct high spatial frequencies in super-resolution tasks, resulting in blurred super-resolved images [30]. We therefore suggest a new loss function, weighted frequencies error (WFE), which takes the mean absolute error (MAE) in the Fourier domain and applies higher weights for higher frequencies, to give more penalty to higher frequencies, so during the training the network will put more effort on reconstruction of higher frequencies. This loss function is defined by:

$$\mathcal{L}_{WFE} = \sum_i \sum_j w_{i,j} \cdot |\mathcal{F}\{I_{generated}\} - \mathcal{F}\{I_{target}\}|. \qquad (1)$$

An illustration of the loss is demonstrated in Figure 2. Note that we did not combine additional losses to our loss function, as by experiment with combined losses such as MAE, MSE, SSIM [31], and perceptual loss [32], we did not achieve better results.

Fuoli et al. [30] suggested using a loss function in the frequency domain for the same problem. However, they split the loss function into two different losses: magnitude and phase. We, on the other hand, suggest one loss, which takes the subtraction of the complex numbers of the predicted and reference images per pixel and then applies L1 norm on it, resulting in one unified loss in the frequency-domain. In addition, we add the spatial-frequency weights to give more penalty for higher frequencies.

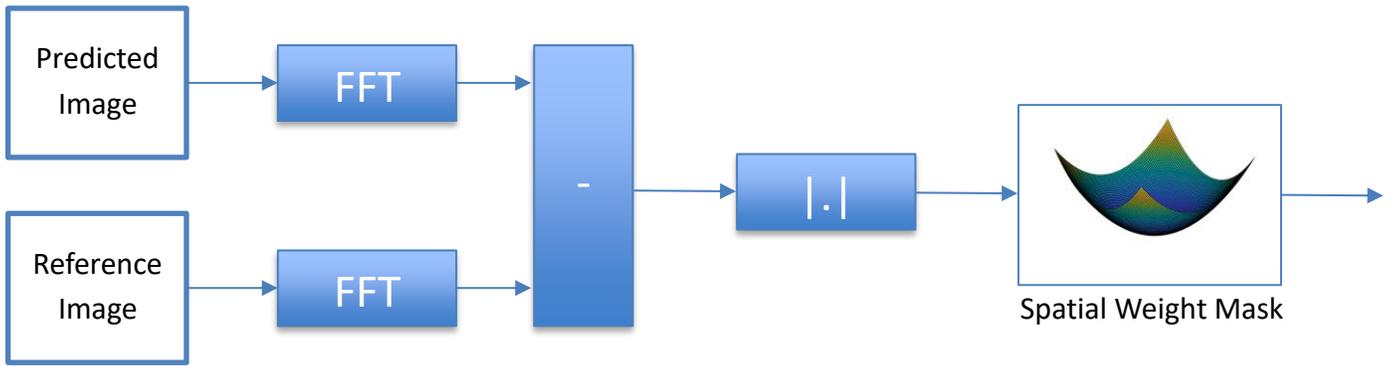

**Figure 2.** Block diagram of the weighted FFT loss function, resulting in the ability to focus on the reproduction of the high frequencies in the image, which is crucial in the task of super resolution.

## Model configuration

For comparison, we tested our method and ESRGAN on x4 and x8 super-resolution tasks. The input to the network is the x4 and x8 down-sampled image of each of the 256x256-pixel RGB frozen-section patches.
We trained our model using batch size of 2 and learning rate of $3 \cdot 10^{-6}$. We used 4 epochs for the task of x4 super-resolution and 6 epochs for the task of x8 super-resolution.
We trained the ESRGAN using batch size of 4 and learning rate of $10^{-4}$. Like in our method, we trained the ESRGAN using 4 epochs for the task of x4 super-resolution and 6 epochs for the task of x8 super-resolution. The ESRGAN has also gradient penalty with weight of 10.
Both methods use Adam as the optimizer.

## Evaluation metrics

As our research aim to eventually serve for medical diagnosis, it is crucial to obtain high performance in distortion metrics rather than perception metrics. In other words, we want to achieve high fidelity images with better correspondence to the real images, rather than sharp and photo-realistic images, which contain fake information that might mislead and provide wrong diagnosis. Therefore, we evaluated our results using SSIM and PSNR, which are distortion metrics, and compared our results with the best results we achieved by training the ESRGAN on the same data.

SSIM measures the correspondence between two images by comparing the luminance, contrast, and structure of them, and is defined by:

$$SSIM(x,y) = \frac{(2\mu_x\mu_y+c_1)(2\sigma_{xy}+c_2)}{(\mu_x^2+\mu_y^2+c_1)(\sigma_x^2+\sigma_y^2+c_2)}, \quad (2)$$

where $(x,y)$ indicate the image pixel location, $\mu_x$ is the pixel intensity mean across $x$, $\mu_y$ is the pixel intensities mean across $y$, $\sigma_x^2$ is the variance of the intensities across $x$, $\sigma_y^2$ is the variance of the intensities across $y$, $\sigma_{xy}$ is the cross-correlation of the pixel intensities across $x$ and $y$, and $c_1, c_2$ are defined by:

$$c_1 = (k_1 L)^2, \; c_2 = (k_2 L)^2, \quad (3)$$

where $L$ is the dynamic range, and $k_1, k_2$ are set to the default values of $k_1 = 0.01$, $k_2 = 0.03$.
The second metric, PSNR, is defined by:

$$PSNR(x,y) = 20 \log_{10} \frac{MAX_I}{\sqrt{MSE(x,y)}}, \quad (4)$$

where:

$$MSE(x,y) = \frac{1}{mn}\sum_{i=0}^{m-1}|x(i,j)-y(i,j)|^2, \quad (5)$$

and $MAX_I$ is the maximum intensity value.

# Results

After training the network, the results were collected on the test data of 2,651 RGB 256x256-pixel frozen-section patches.
Figures 3 and 4 present a few examples of our results compared to bicubic interpolation and ESRGAN. In the x8 super-resolution task, our method results in better correspondence to the high-resolution image. As mentioned earlier, this fits our requirement to prefer higher correspondence over photo-realistic image with poor correspondence that might lead to wrong diagnosis.

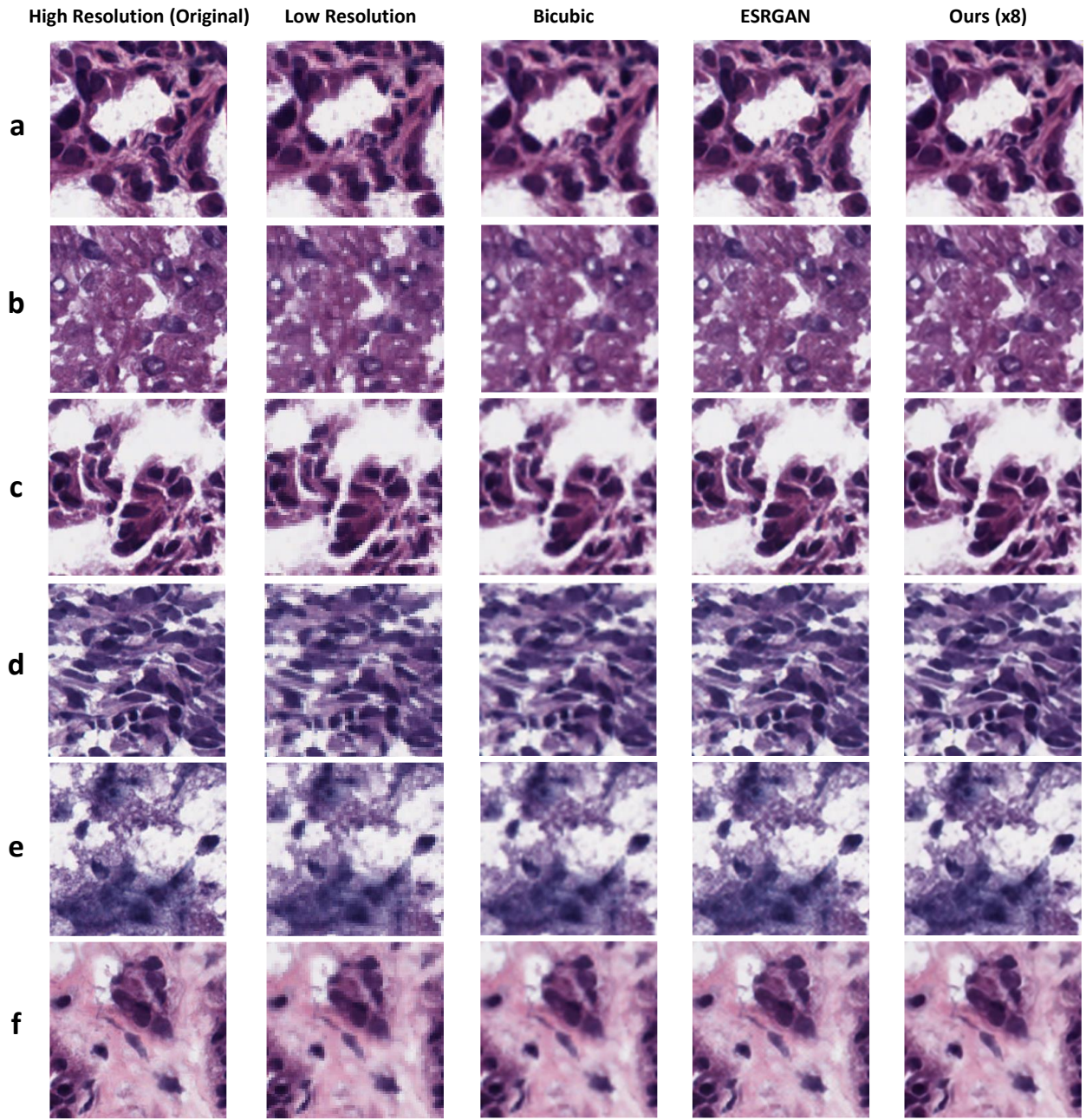

**Figure 3.** Example of **x4** super resolution task on frozen-section patches. Comparison of ESRGAN and our method for each image (SSIM/PSNR): (a) ESRGAN – 0.89/30.18, ours – 0.91/31.1, (b) ESRGAN – 0.82/28.58, ours – 0.84/29.5, (c) ESRGAN – 0.90/29.98, ours – 0.91/30.74, (d) ESRGAN – 0.87/28.64, ours – 0.89/29.79, (e) ESRGAN – 0.84/28.85, ours – 0.86/29.85, (f) ESRGAN – 0.86/31.68, ours – 0.88/33.04.

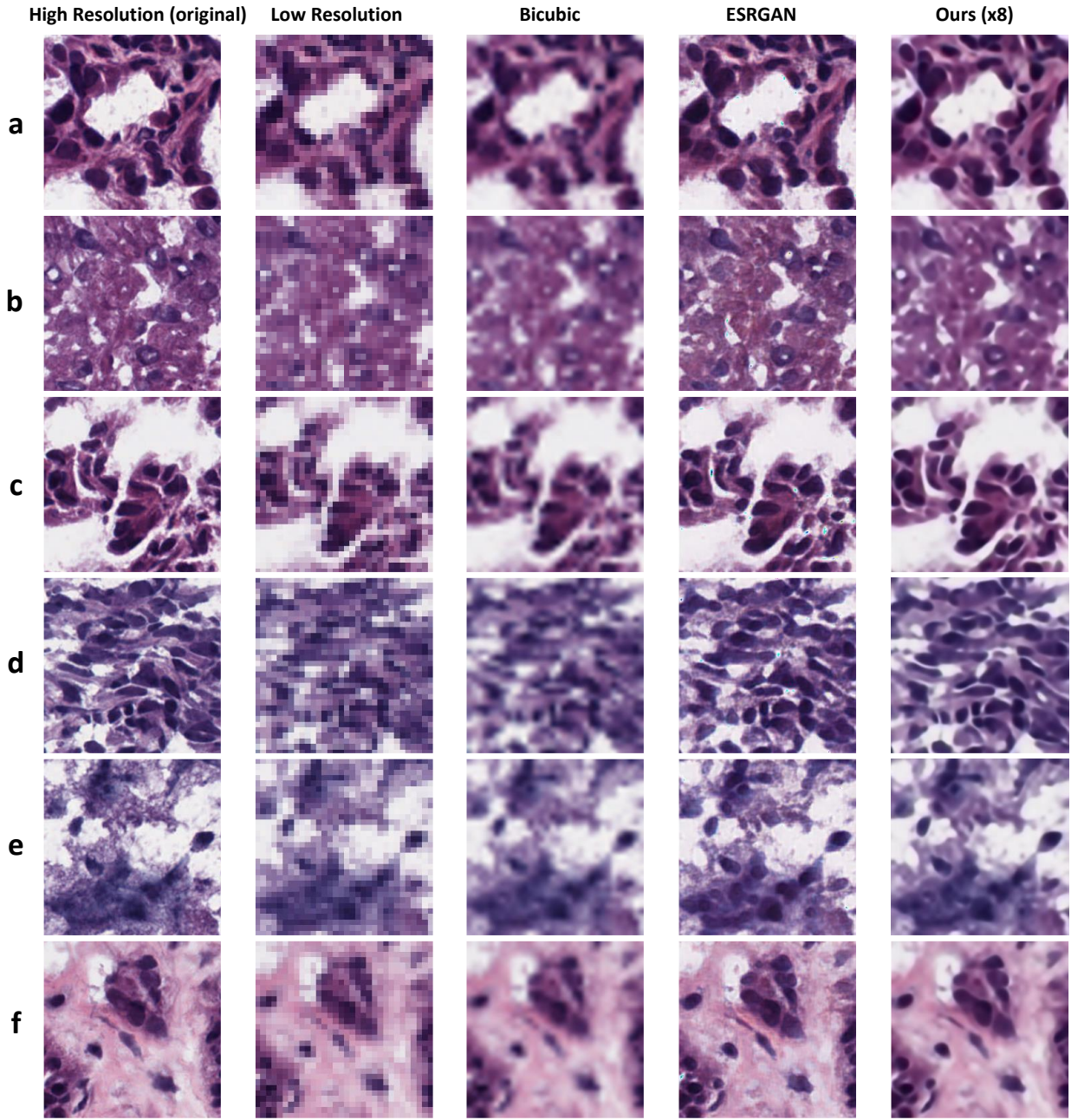

**Figure 4.** Example of **x8** super resolution task on frozen-section patches. Comparison of ESRGAN and our method for each image (SSIM/PSNR): (a) ESRGAN – 0.65/23.07, ours – 0.74/25.01, (b) ESRGAN – 0.48/21.55, ours – 0.63/24.14, (c) ESRGAN – 0.70/22.93, ours – 0.78/25.09, (d) ESRGAN – 0.58/21.01, ours – 0.69/23.47, (e) ESRGAN – 0.55/22.25, ours – 0.66/24.64, (f) ESRGAN – 0.62/25.08, ours – 0.75/28.15.

As presented in Table 1, for x4 super-resolution task, our method is leading with slightly better results in terms on average MSE, PSNR and SSIM. For x8 super-resolution task, our method outperforms with significant improvement in terms of these metrics. In case of frozen-section histology, the inference time of the super-resolution model is critical, as the target is to scan faster to provide rapid frozen-section imaging during surgeries. We measured the inference time of each method on an ASUS TUF

Dash F15 i7 PC, with NVIDIA GeForce RTX 3060 GPU. As can be seen from Table 1, for both x4 and x8 super-resolution tasks the inference time of our method is 3-4 times faster than the ESRGAN. Note that we did not compare our results to diffusion models since the inference time of the diffusion models is much slower and currently not suitable for rapid imaging.

| Method | Upscale Task | MSE | PSNR [dB] | SSIM | Inference Time Per Patch [ms] |
|---|---|---|---|---|---|
| Bicubic | x8 | $533 \cdot 10^{-5}$ | 22.939 | 0.651 | - |
| ESRGAN | x8 | $603 \cdot 10^{-5}$ | 22.397 | 0.62 | 46.8 |
| Ours | x8 | $\mathbf{313 \cdot 10^{-5}}$ | **25.296** | **0.73** | **13.5** |
| Bicubic | x4 | $145 \cdot 10^{-5}$ | 28.633 | 0.85 | - |
| ESRGAN | x4 | $107 \cdot 10^{-5}$ | 29.926 | 0.878 | 75.4 |
| Ours | x4 | $\mathbf{86 \cdot 10^{-5}}$ | **30.911** | **0.8936** | **16.9** |

**Table 1.** Results of our method compared to bicubic interpolation and ESRGAN. As can be seen from the table our method outperforms in terms of MSE, PSNR, SSIM (averaged on the entire test set) and inference time for both x4 and x8 super-resolution tasks.

Another advantage in our method is the model convergence during training. As our model is based on the bilinear interpolation residual as a baseline, its effort is to correct the interpolation error rather than performing full image reconstruction. This leads to a better performance in the initial point and during the training convergence. Figure 5 demonstrates this idea, as a comparison between our method to ESRGAN in terms of model convergence to the SSIM metric during training.

Note that also in terms of time per epoch we get better results using our method. For example, for the x8 super-resolution task the average training time of our model takes 43 minutes per epoch while for the ESRGAN it is 83 minutes per epoch.

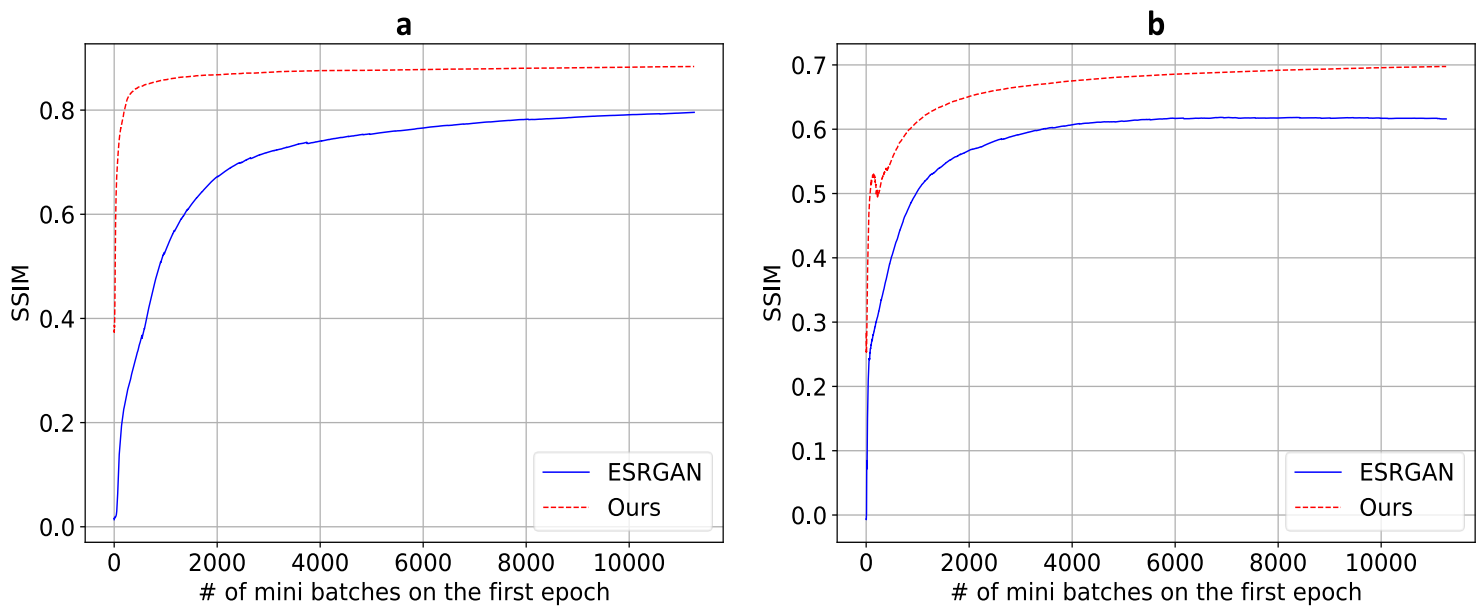

**Figure 5.** Comparison of the SSIM metric during training between our method and the ESRGAN. It can be seen from the graphs that our method has better initial condition and convergence in terms of SSIM for the super-resolution tasks of (a) x4 and (b) x8.

Note that our model was trained only on kidney data. To test the generalization of our method for other organ biopsies, we tested our model on test data of breast biopsy patches. The test data has 1,050 patches of breast frozen-section biopsies. For the task of x4 super-resolution, we obtained an average SSIM of 0.88 and an average PSNR of 32.12, while the ESRGAN obtained an average SSIM of 0.87 and an average PSNR of 31.28. For the task of x8 super-resolution, we obtained an average SSIM of 0.74 and an average PSNR of 27.11, while the ESRGAN obtained an average SSIM of 0.63 and an average PSNR of 24.03, surprisingly showing even slightly better results in comparison to the kidney test data. Examples of x4 and x8 super-resolution tasks on breast data can be seen in Figures 6 and 7.

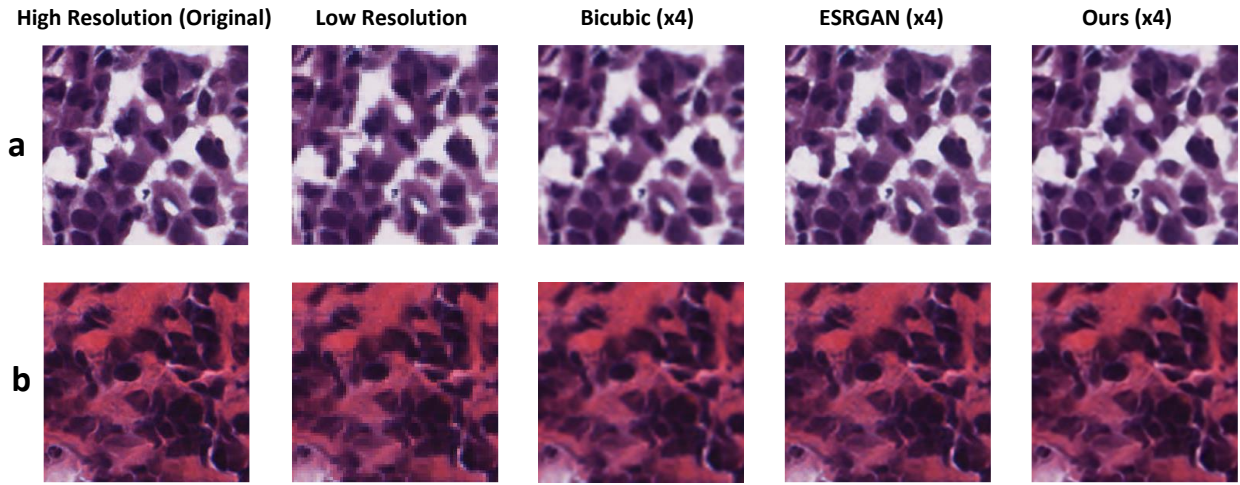

**Figure 6.** Example of **x4** super resolution task on frozen-section patches of breast. Comparison of ESRGAN and our method for each image (SSIM/PSNR): (a) ESRGAN – 0.89/30.68, ours – 0.90/31.44, (b) ESRGAN – 0.89/32.22, ours – 0.89/32.02.

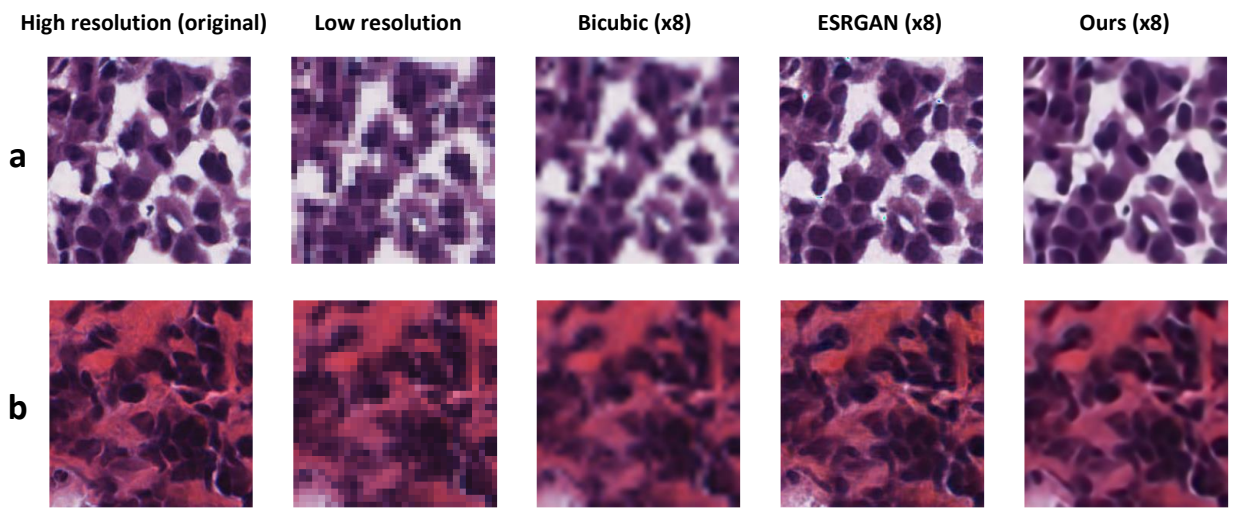

**Figure 7.** Example of **x8** super resolution task on frozen-section patches of breast. Comparison of ESRGAN and our method for each image (SSIM/PSNR): (a) ESRGAN – 0.70/23.76, ours – 0.80/26.23, (b) ESRGAN – 0.68/ 25.18, ours – 0.77/ 27.13.

Finally, to assess the enhancement in manual diagnostic capabilities facilitated by our super-resolution model, we conducted a clinical pathologist's evaluation using stitched images generated from a grid of 4 by 4 sub-regions extracted from low-resolution 32x32-pixel RGB frozen sections alongside their corresponding super-resolution counterparts. A qualified pathologist (G.G.) performed the tissue analysis based on both the low-resolution and super-resolution images. The pathologist's evaluation revealed noteworthy disparities between the two image types. Specifically, when examining the low-resolution images, critical cytological aspects, such as distinguishing the nucleus from the cytoplasm and delineating cell boundaries, proved to be unfeasible in the low-resolution images. For instance, as illustrated in Figure 8, a region marked by a red circle displayed a cluster of cells. The inherent high cellularity of the tissue and the absence of uniform cell orientation rendered cell boundary delineation impossible in the low-resolution image. Additionally, in the region demarcated by a green circle in Figure 8, accurate assessment of cell orientation was impeded in the low-resolution image. In Figure 9, the region indicated by a red circle exhibited high cellularity, making it impossible to discern the nucleus accurately in the low-resolution images, consequently hindering the assessment of cytological atypia. Furthermore, distinguishing neoplastic cells from immune cells in the lower resolution images proved challenging. This differentiation is a pivotal component of most frozen section biopsy evaluations, as it allows for the discrimination between neoplastic and inflammatory processes, underscoring the clinical significance of our super-resolution model's advancements in pathological diagnosis.

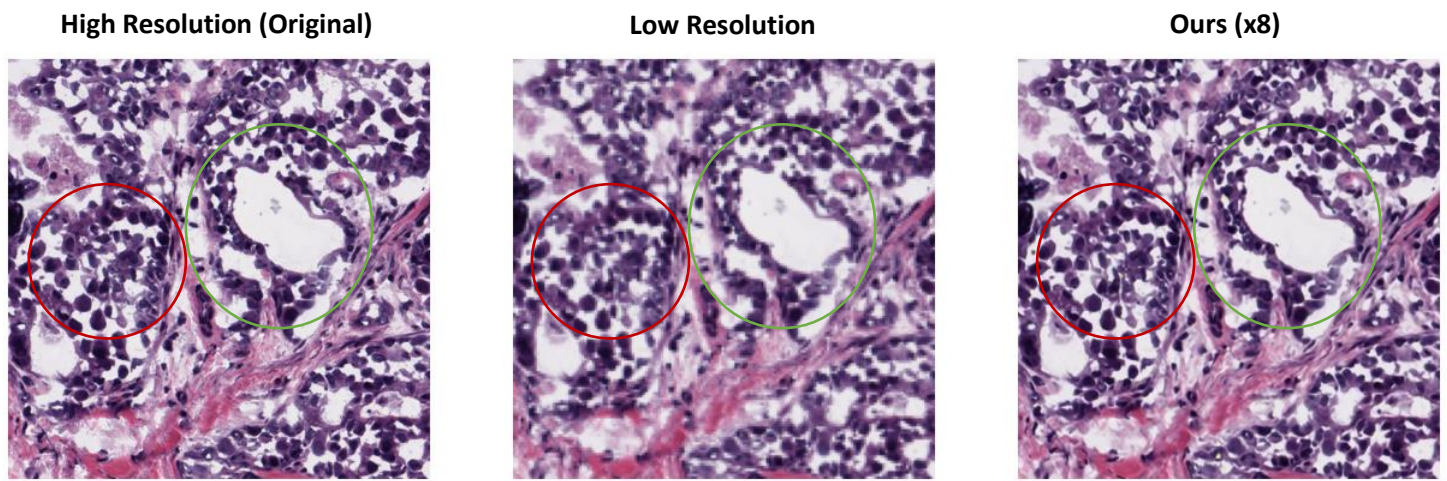

**Figure 8.** First example of pathologist's analysis: Looking at low-resolution frozen-section image (middle image), it is not possible to assess cytologic atypia, separating the nucleus from the cytoplasm and defining the borders of each cell. The area marked by a red circle shows a cluster of cells, where due to the high cellularity of the tissue and lack of uniform cells orientation, it is not possible to define the borders of the cells in the low-resolution image. In the area marked by a green circle, the cell orientation cannot be accessed correctly in the low-resolution image. This is solved in the resolution increase model (right image).

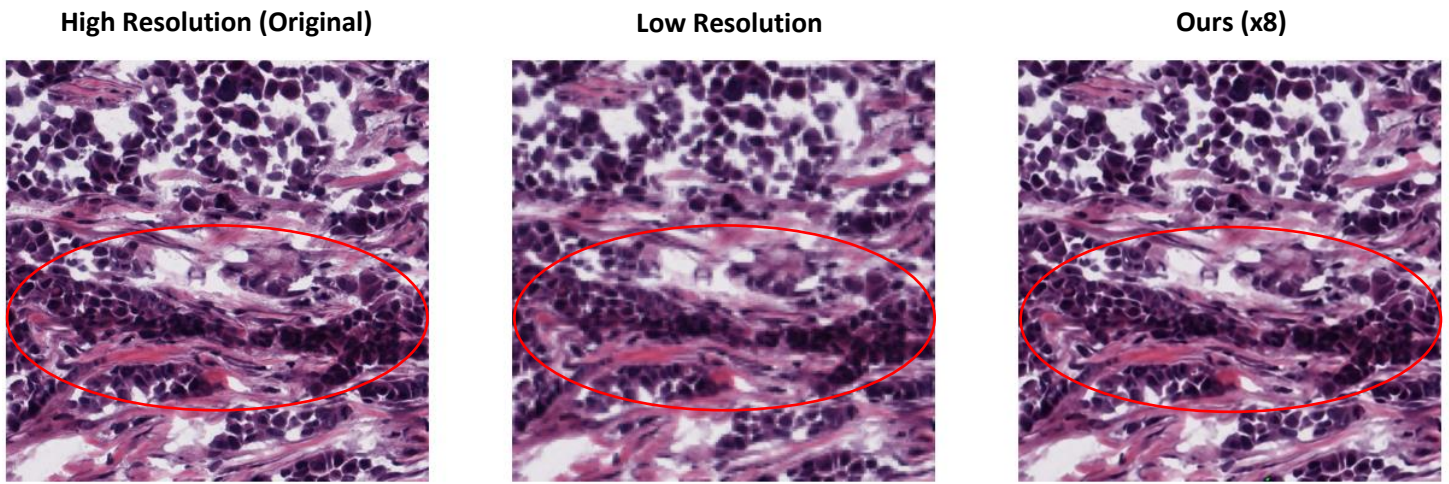

**Figure 9.** Second example of pathologist's analysis: Looking at low-resolution frozen-section image (middle image), it is not possible to assess the nucleus properly. The region marked by red circle is of high cellularity; therefore, the level of atypia cannot be assessed. Also, it is difficult to differ neoplastic cells from immune cells in the lower resolution images. This is solved in the resolution increase model (right image).

# Discussion

We have presented a new method for obtaining super-resolution of frozen-section biopsies, being able to obtain better SSIM, PSNR and inference time compared to existing methods. We prioritize distortion metrics over perception metrics due to the inherent risk associated with visually sharp and photo-realistic images that lack correspondence with ground-truth data, potentially leading to erroneous diagnoses.

After demonstrating x4 resolution improvement, we transited into the more intricate domain of x8 super-resolution, where the inherent advantages associated with our method become more pronounced. Despite achieving commendable results in terms of distortion metrics, our images exhibit some blurriness. This degradation in image quality arises from the necessity to replace one intensity for each RGB pixel with 64 intensity values, which can be not achievable due to lack of information in the low-resolution image.

Since the true quality metric for our method is the improved diagnostic ability of the pathologist with the super-resolved images, future investigations involving more pathologists is mandatory. Furthermore, exploring the implementation of secondary algorithms for image sharpening, followed by a reevaluation of the associated metrics to evaluate and test the effectiveness of such post-processing techniques, are expected to further improve the results.

We deliberately selected ESRGAN as the benchmark rather than diffusion models or transformer-based architectures, due to their slower inference time. Our primary objective was to achieve a balance between computational efficiency, minimizing hardware requirements, and preserving high-resolution image fidelity. Diffusion models and transformer-based architectures often entail significantly longer inference times and may also, like ESRGAN, compromise perception metrics in favor of distortion metrics.

In summary, our study presents a new approach for achieving robust super-resolution in the medical domain, and specifically for frozen sections biopsies, where diagnostic decisions hinge on the quality of these images. Our contribution introduces a novel architecture that leverages bilinear interpolation as a residual baseline, followed with an attention U-net to correct interpolation errors. Additionally, we propose the use of a weighted-frequency loss function to push the network to allocate more effort toward reconstructing high-frequency components, which are inherently challenging in super-resolution tasks.

# Supporting information

Not applicable.

# Conflict of interest

The authors declare no conflict of interest.

# Code availability

The code and the dataset reference are available upon request.

# Data availability statement

The slide images we used to train and test our models were taken from the TCGA Research Network: [https://www.cancer.gov/tcga](https://www.cancer.gov/tcga)

# Keywords

super-resolution, deep learning, histopathology, attention, frequency loss, histopathology, frozen section